\documentclass[british,review,number]{article}
\usepackage[T1]{fontenc}
\usepackage[latin9]{inputenc}
\usepackage{geometry}
\usepackage{array}
\usepackage{booktabs}
\usepackage{multirow}
\usepackage{amsmath}
\usepackage{amssymb}
\usepackage{graphicx}
\PassOptionsToPackage{version=3}{mhchem}
\usepackage{mhchem}

\makeatletter

\usepackage{siunitx}

\makeatother

\usepackage{babel}
\begin{document}

\title{Genetic algorithms and solid state NMR pulse sequences}

\author{Matthias Bechmann\thanks{matthias.bechmann@york.ac.uk}\\
	Department of Chemistry\\
	University of York\\
	YO10 5DD York, UK\\
	\and John Clark\thanks{john@cs.york.ac.uk}\\
	Department of Computer Science\\
	University of York\\
	YO10 5DD York, UK\\
	\and Angelika Sebald\thanks{angelika.sebald@york.ac.uk}\\
	Department of Chemistry\\
	University of York\\
	YO10 5DD York, UK}
\date{17th December 2012}
\maketitle

\begin{abstract}
The use of genetic algorithms for the optimisation of magic angle
spinning NMR pulse sequences is discussed. The discussion uses as
an example the optimisation of the $\mathrm{C}7_{2}^{1}$ dipolar
recoupling pulse sequence, aiming to achieve improved efficiency for
spin systems characterised by large chemical shielding anisotropies
and/or small dipolar coupling interactions. The optimised pulse sequence
is found to be robust over a wide range of parameters, requires only
minimal \textit{a priori} knowledge of the spin system for experimental
implementations with buildup rates being solely determined by the
magnitude of the dipolar coupling interaction, but is found to be
less broadbanded than the original $\mathrm{C}7_{2}^{1}$ pulse sequence.
The optimised pulse sequence breaks the synchronicity between r.f.
pulses and sample spinning.
\end{abstract}

\section{Introduction}

Solid state magic angle spinning (MAS) NMR spectroscopy has become
an indispensable and rather widely used tool for the characterisation
of crystalline and non-crystalline powder materials. Amongst the magnetic
interactions present, direct dipolar coupling plays a particularly
prominent role owing to its direct relationship with internuclear
distances ($\propto r^{-3}$), making the measurement of direct dipolar
coupling constants a highly attractive target from a structural point
of view. In addition, NMR\ experiments employing cross correlations,
multi-quantum excitation or polarisation transfer \cite{Spiess2007,Laws2002a}
rely on the presence of direct dipolar coupling. Unsurprisingly, over
the years much effort has been devoted to the development of MAS NMR
pulse sequences that make direct dipolar coupling information accessible
in an accurate, quantifiable and straightforward fashion \cite{Jaroniec2000,Brinkmann2001,Chen2010,Carravetta2001,Brinkmann2002,Guenne2003,Brouwer2005}.

The theoretical description and development of MAS NMR experiments
is typically accomplished by using the frameworks of average Hamiltonian
(AHT) \cite{Waugh2007} or Floquet \cite{Leskes2010} theory. Application
of these theories is difficult if the interaction of interest is not
the dominating one or, as is the case of homonuclear direct dipolar
coupling, renders the system Hamiltonian homogeneous. In such circumstances
the description of the spin dynamic is only possible approximately.
Consequently, pulse experiments derived by using such approximations
are starting to fail if e.g. direct dipolar coupling is not the strongest
interaction (long internuclear distances) and/or if other interactions,
especially chemical shielding anisotropy (particularly at high external
magnetic field strengths $\boldsymbol{B}_{0}$) are dominant. MAS
NMR experiments that do not conform to the assumptions of lower-order
AHT and Floquet theories can sometimes be improved by either using
higher order approximations \cite{Hohwy1997,Brinkmann2004}, or by
combinations of supercycling \cite{Levitt2002a,Kristiansen2004,Kristiansen2006}
and composite pulses \cite{Levitt1996}. Furthermore, owing to the
nature of these theoretical tools, MAS NMR pulse sequences typically
consist of rotation-synchronised r.f. pulse trains. While the assumption
of rotation synchronisation greatly facilitates the mathematical description,
it excludes any non-synchronous experiments.

As an alternative approach numerical simulations of MAS NMR spin dynamics
combined with various search algorithms have been used to improve
existing experiments or to find new or improved pulse sequences. This
approach has been made feasible by easy access to increasing computing
power and advances in spectrometer hardware technology. Successful
transfer of optimised numerical results to a real-world NMR spectrometer
heavily relies on robust and finely tuneable hardware parameters.
Such approaches have resulted in, for example, techniques such as
strongly modulating pulses \cite{Fortunato2002,Boulant2008,Mitra2008},
a range of modified pulse sequences based on optimal control theory
\cite{Glaser1998,Tosner2009,Nielsen2009}, or modulated decoupling
schemes \cite{Sakellariou2000,Paepe2003}. All these approaches have
in common that they employ classical optimisation techniques based
on Simplex- or Newton-type methods \cite{Chong2008} which exploit
the local shape of a particular fitness function in order to maximise
performance. Numerical simulation-based optimisation approaches have
also occasionally taken advantage of genetic algorithms (GA) instead
of relying on classical optimisation \cite{Freeman1987,Gray2000,Pang2007,Herbst2009b,Grimminck2011}.

Here we discuss that GAs offer a class of stochastic search algorithms
that is able to explore parameter spaces more widely than classical
algorithms and, hence, may be more suitable to find unconventional
pulse sequences not easily accessible by other means. In particular,
constraints can be chosen such that resulting pulse sequences are
straightforward, enabling insight into the spin dynamics and, hence,
offering a new starting point for the further development of theoretical
descriptions. We describe the necessary requirements to use a GA in
the search for improved MAS NMR experiments and apply this to a representative
test case, the $\mathrm{C}7_{2}^{1}$ pulse sequence \cite{Levitt2002a}
in the presence of large chemical shielding anisotropies (CSA) and
small dipolar coupling interactions. The results are analysed with
regard to robustness, possible trends and experimental verification.
The properties and performance of the GA are compared to that of other
(classical) search algorithms.

\section{Results and Discussions}

NMR experiments and pulse sequences in particular depend on multiple
parameters and tend to have a modular structure. MAS NMR pulse sequences
often display periodic repetition of basic r.f. pulse elements and
the variation of these elements by modulation of a single parameter
(commonly a r.f. pulse phase). This internal structure originates
from the response of nuclear magnetic moments to external magnetic
fields or sample rotation, and from the way theory is describing the
spin response to these external perturbations \cite{Waugh2007,Leskes2010},
often explicitly involving symmetry features \cite{Levitt2002a}.
Any numerical approach to pulse sequence design and/or optimisation
needs to make the decision to either retain (some of) these symmetry
principles or to ignore internal symmetry and structure in favour
of an unbiased search approach. The first approach can reduce the
dimension of the parameter space and results are likely to be closely
linked to the underlying theoretical description of the spin dynamics
but will require, to some extent, previous knowledge of the spin dynamics.
The second approach is typically characterised by a potentially larger
parameter space and a more complex structure with limited potential
to provide insight into the spin dynamics. A hybrid approach between
this two extremes is, for example, the unbiased optimisation of basic
NMR pulses ($\SI{90}{\degree}$, $\SI{180}{\degree}$, ...) to compensate
for experimental imperfections while keeping the overall pulse sequence
unaltered: while making the pulse sequence experimentally more robust,
this would not have the potential to improve the performance of the
experiment in general.

Analysing the experimental performance of pulse sequences as a function
of pulse sequence parameters by numerical simulations \cite{Bjerring2003,Marin-Montesinos2005,Leskes2011}
commonly yields rugged error surfaces with multiple local minima,
sharp and often singular resonance conditions, and areas of rather
low variation inbetween. The exploration of such error landscapes
in the search for new or improved pulse sequences is a difficult task
for classical optimisation routines if the search starts far away
from the global minimum, and if no conjectures about a likely location
of optimum parameters can be made in advance. Stochastic search algorithms
such as evolutionary and genetic algorithms (GA) \cite{Hibbert1993,Schwefel1993,Holland1975}
are much better placed to deal with this type of search scenario for
a number of reasons. GAs provide high degrees of flexibility regarding
the choice of fitness functions as compared with e.g. least-squares
methods. GAs are independent of gradients with concomitant gains in
speed especially if no analytically defined gradients exist. GAs can
handle mixed parameters (such as continuous and discrete parameters)
naturally well. All these features make the application of GAs an
attractive alternative for purposes of NMR pulse sequence searches
and optimisations.

In the following we demonstrate the application of a GA to the optimisation
of a homonuclear dipolar recoupling MAS NMR pulse sequence, $\mathrm{C7_{2}^{1}}$
\cite{Levitt2002a}. We use this pulse sequence as an example because
it is widely used \cite{Levitt2002a} and its practical advantages
and disadvantages are thus well documented. Our objective is the optimisation
of $\mathrm{C7_{2}^{1}}$ performance for spin systems characterised
by the simultaneous presence of large CSA and small dipolar coupling
constants such that internuclear distances can be determined - a task
not easily achieved by $\mathrm{C7_{2}^{1}}$ in its original or some
of its modified \cite{Carravetta2000,Karlsson2003} forms.

\subsection{GA optimisation of the $\mathrm{C}7_{2}^{1}$ pulse sequence}

\begin{figure}
\begin{centering}
\includegraphics[width=0.6\columnwidth]{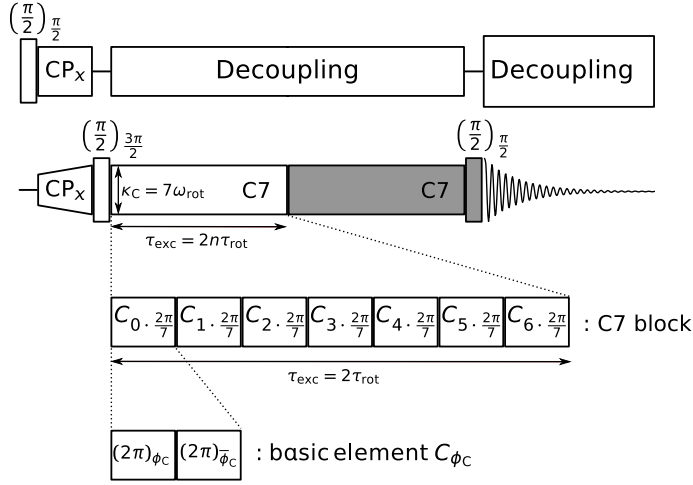}
\par\end{centering}

\caption{\label{fig:C7-DQF-pulse-sequence}The $\mathrm{C}N_{n}^{\nu}=\mathrm{C}7_{2}^{1}$
pulse sequence for excitation and reconversion of double-quantum coherences
\cite{Levitt2002a}}
\end{figure}

The $\mathrm{C7_{2}^{1}}$ pulse sequence (see Figure \ref{fig:C7-DQF-pulse-sequence})
 can be parametrised as 
\begin{equation}
\mathrm{C}N{}_{n}^{\nu}=\left\{ \left[\left(\theta_{\mathrm{C}}\right)_{\mathrm{\phi_{\mathrm{C}}}}\left(\theta_{\mathrm{C}}\right)_{\mathrm{\overline{\phi}_{\mathrm{C}}}}\right]_{\phi_{\mathrm{C}N}}^{N}\right\} ^{n_{\mathrm{C}}}\label{eq:C7parametrisation}
\end{equation}
with the space and spin winding number $n=2$ and $\nu=1$, the basic
element $\mathrm{C}=\left(\theta_{\mathrm{C}}\right)_{\mathrm{\phi_{\mathrm{C}}}}\left(\theta_{\mathrm{C}}\right)_{\mathrm{\overline{\phi}_{\mathrm{C}}}}$,
pulse flip angle $\theta_{\mathrm{C}}=2\pi$, pulse phases $\phi_{\mathrm{C}}=0$
and $\overline{\phi}_{\mathrm{C}}=180$, $N=7$ iterations of the
$\mathrm{C}$ elements, phase increment $\phi_{\mathrm{C}N}$ of the
$\mathrm{C}$ elements, and $n_{c}$ iterations of $\mathrm{C7}$
r.f. blocks. With pulse amplitude $\kappa_{\mathrm{C}}$ and duration
$\tau_{\mathrm{C}}$, symmetry parameters $N$, $n$ and $\nu$ \cite{Levitt2002a}
demand the relations 
\begin{eqnarray}
\kappa_{\mathrm{C}} & = & \frac{2N}{n}\omega_{\mathrm{rot}}=7\cdot\omega_{\mathrm{rot}}\\
\kappa_{\mathrm{C}}\tau_{\mathrm{C}} & = & \theta_{\mathrm{C}}\\
\phi_{\mathrm{C}N} & = & \frac{2\pi\cdot\nu}{N}
\end{eqnarray}
For a given spinning speed $\omega_{\mathrm{rot}}$ this parametrisation
fixes all possible parameter values. For the optimisation we now keep
the seven-step phase increment $\phi_{\mathrm{C}N}=\phi_{\mathrm{C7}}$
and the number of pulses per $\mathrm{C7}$ block ($14=7\cdot2$). All
other constraints are removed and $\mathrm{\kappa_{\mathrm{C}}\rightarrow}\kappa_{1},\kappa_{2}$,
$\mathrm{\tau_{\mathrm{C}}\rightarrow}\tau_{1},\tau_{2}$,$\mathrm{\phi_{\mathrm{C}}\rightarrow}\phi_{1},\phi_{2}$
giving a pulse sequence parametrisation 
\begin{equation}
\mathrm{C}7{}_{2}^{1}\mathrm{_{opt}}=\left\{ \left[\left(\kappa_{1}\tau_{1}\right)_{\mathrm{\phi_{\mathrm{1}}}}\left(\kappa_{2}\tau_{2}\right)_{\mathrm{\phi_{2}}}\right]_{\phi_{\mathrm{C7}}}^{7}\right\} ^{n_{\mathrm{C}}}\label{eq:C7opt}
\end{equation}
with seven variable parameters ($\kappa_{1}$, $\kappa_{2}$, $\tau_{1}$,
$\tau_{2}$, $\phi_{1}$, $\phi_{2}$, $n_{\mathrm{C}}$).

Double-quantum filtration (DQF) experiments are constructed using
Equation \ref{eq:C7opt} in the usual way (see Figure \ref{fig:C7-DQF-pulse-sequence}) 

\begin{equation}
FID=\mathrm{C}7{}_{2}^{1}\mathrm{_{opt}}-\mathrm{DQF-\left(\mathrm{C}7{}_{2}^{1}\mathrm{_{opt}}\right)_{90}}-\mathrm{acq}\label{eq:DQF-FID}
\end{equation}
with a $\SI{90}{\degree}$ r.f. phase shift during reconversion to
achieve absorptive lineshapes. This pulse-sequence template with seven
variable parameters is now interfaced with the GA to generate candidate
solutions for improved NMR experiments.

\begin{figure}
\begin{centering}
\includegraphics[width=0.4\columnwidth]{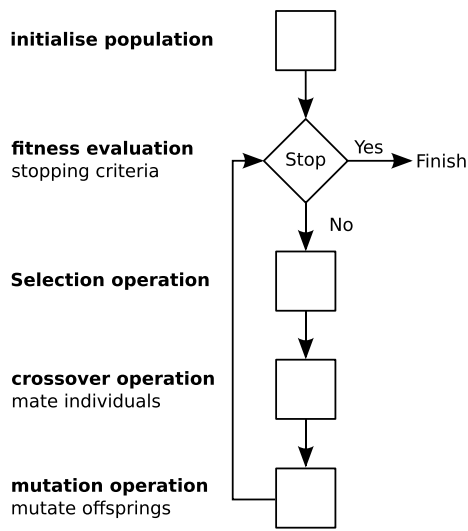}
\par\end{centering}

\caption{\label{fig:Flow-chart-GA}Flowchart of the procedure of a genetic
algorithm.}
\end{figure}

The GA simulates the selection from a population of candidate solutions
with the goal of evolving the population to a more \emph{fit} and
more \emph{diverse} next generation. This evolution process is accomplished
by applying genetic operators like inheritance, mutation, selection
and crossover to an existing population of candidate solutions. In
order to use a GA it is necessary to represent the solution to a given
problem (phenotype) as a \emph{genome} or \emph{chromosome }(genotype)\emph{
}\cite{Goldberg1989}\emph{. }Overall every genetic algorithms requires
the implementation of three crucial components: 1) an objective/fitness/cost
function, 2) the definition and implementation of the genetic representation/encoding
of the problem description and 3) the definition and implementation
of the genetic operators: 
\begin{enumerate}
\item The objective function $f$ is calculated from Equation \ref{eq:DQF-FID}
as 
\begin{equation}
f=1-FID\left(1\right)\label{eq:fitness}
\end{equation}
This takes the normalised, first-point intensity in the $FID$ as
representative of the whole integrated spectral area and hence the
DQF efficiency of the pulse sequence. The normalised spectral intensity
$FID(1)$ can assume positive and negative values $[-1,1]$ in homonuclear
recoupling DQF experiments \cite{Brinkmann2000}, but for the operation
of the algorithm it is advantageous to only have positive fitness
values. This is achieved by the form of Equation \ref{eq:fitness}
where more fit candidates have smaller values $f$. Experimentally,
this choice of fitness function means that here we concentrate on
improved efficiency of the pulse sequence. In some instances 
the single-quantum elements  in the final density matrix can 
be an alternative measure of DQF efficiency. 

\item The parameters in Equation \ref{eq:C7opt} control the performance
of the NMR pulse experiment. Therefore, these parameters are the real-world
parameters of our search problem and are referred to as phenotype
of the problem description. GA operators do not work on this description,
but on its genetic encoding, the genotype. This genetic representation
of the pulse sequence is achieved by mapping each pulse-sequence parameter
to a bit-string representation, the genes. Together these genes form
the genome/chromosome of the search problem. The bit-string mapping
is achieved by defining decimal boundary values and a bit depth which
controls the range and resolution over which the parameter can vary
during the optimisation process (Figure \ref{fig:Binary-encoding}).
A suitable choice of bit depth in relation to the boundary values
allows to control encoding of integer and float pulse-sequence parameters.
\item Different pulse sequences correspond to different genomes and together
form a population of a given size. Every member of this population
is a candidate solution to the search problem. A population is evolved
to the next generation by three genetic operations. Individuals/parents
are selected for later crossover/mating using roulette-wheel selection,
also called fitness-proportionate selection. Individuals are more
likely to be selected the fitter they are according to their objective
function. Elitism \cite{Wall1996} is applied to avoid that the stochastic selection
process misses the best candidate of a population. The crossover probability
$p_{\mathrm{c}}$ controls how many individuals of a population are
selected to mate. Here one-point crossover is used as shown in Figure
\ref{fig:One-point-crossover}. Mutation is the process of changing
elements in the genome by a random process. A random flip mutator
is used that toggles the bit value in a gene according to a mutation
probability $p_{\mathrm{m}}$ (Figure \ref{fig:Flip-mutation}).
\begin{figure}
\begin{centering}
\includegraphics[width=0.5\columnwidth]{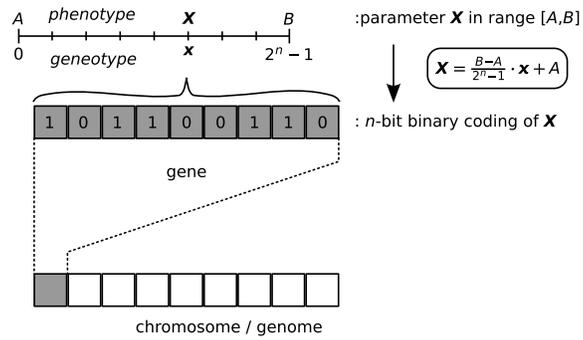}
\par\end{centering}

\caption{\label{fig:Binary-encoding}Pulse sequence parameter $\boldsymbol{X}$
is variable over interval $\left[A,B\right]$. It is binary encoded
as a bit-string $\boldsymbol{x}$ of size $n$ to form a gene. Together
with the other encoded parameters the genes form the genome or chromosome
of our problem.}
\end{figure}
\begin{figure}
\begin{centering}
\includegraphics[width=0.5\columnwidth]{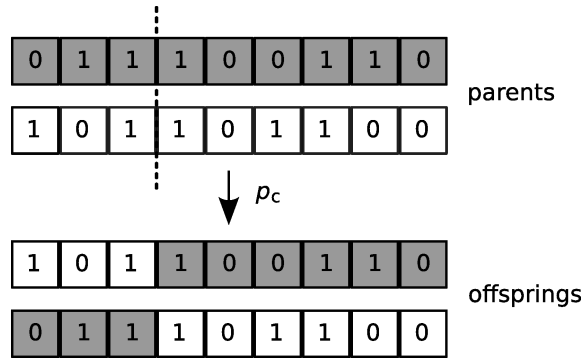}
\par\end{centering}

\caption{\label{fig:One-point-crossover}One-point crossover at a random position.}
\end{figure}
\begin{figure}
\begin{centering}
\includegraphics[width=0.3\columnwidth]{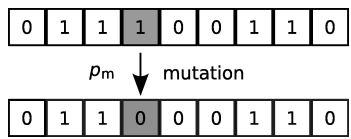}
\par\end{centering}

\caption{\label{fig:Flip-mutation}Flip mutation genetic operator.}
\end{figure}

\end{enumerate}
Various GAs can be created by the way the components 1--3 are implemented
and combined. The overall operation of the GA follows the steps shown
in Figure \ref{fig:Flow-chart-GA}.
\begin{itemize}
\item An initial population of $N_{\mathrm{P}}$ pulse sequences is generated
with random pulse sequence parameter values from within the allowed
parameter ranges. The random assignment of parameter values in the
initial population ensures a diverse distribution of pulse sequences
over the complete search space.
\item The fitness function value $f$ of every pulse sequence is calculated
by simulating the respective NMR experiment. The fitness values represent
point intensities in the overall search space. Individual pulse sequences
are selected for mating based on their fitness value, and deemed unfit
pulse sequences are discarded. This generates a subpopulation of very
fit individuals but also reduces diversity amongst the pulse sequences.
\item Pairs of this subpopulation mate via the crossover genetic operation
until a new population  of pulse sequences is generated.
\item Members of this new population undergo the mutation operation and
afterwards represent the final state of a new generation of pulse
sequences. Mutation is the mechanism that can increase diversity by
being able to change one genome to any other possible genome. Dependent
on how frequently mutation occurs, this drives exploration of the
search space.
\item Fitness evaluation is used to decide on how to progress in the algorithm.
Either the current population is evolved to a new generation, or the
algorithm stops. Stopping criteria can be defined very flexibly and
are typically based on optimum fitness and/or a predefined number
of generations.
\end{itemize}

The performance of a GA not only depends on the implementation of
the genetic operators, but is also susceptible to the tuning of the
parameters: population size $N_{\mathrm{p}}$, crossover probability
$p_{\mathrm{c}}$, mutation probability $p_{\mathrm{m}}$, and the
definition of the fitness function \cite{Goldberg2002}. Their optimum
settings are dependent on the goal of the search task at hand. In
the context of finding new and improved NMR pulse sequences one would
usually like a GA behaviour that explores the parameter search space
widely during the early generations of the algorithm run. This gives
a good chance of collecting a set of diverse and fit pulse sequences.
Later in the run, more localised optimisation towards improving the
best of these fit candidates is preferred.

\subsection{Implementation and exploration of GA optimisation of $\mathrm{C7_{2}^{1}}$}

The performance of GA-generated $\mathrm{C7_{2}^{1}}$-derived pulse
sequences is tested by using the spin system parameters of 1,4$\ce{^{13}C2}$-mono-ammonium
maleate, \textbf{1} (see Figure \ref{fig:1,4--mono-ammonium-maleate-spins}) as input to the numerical spin-dynamics simulations
\cite{Bak2000}. The $\ce{^{13}C}$ spin pair in \textbf{1} displays
a large CSA, no isotropic chemical shielding difference, and a relatively
small dipolar coupling constant $b_{14}/2\pi=\SI{-216}{\hertz}$.
These are characteristics that typically hinder the $\mathrm{C7_{2}^{1}}$
pulse sequence from achieving an optimum DQF efficiency of $\approx73$
percent and lead to DQF-buildup rates that do not solely dependent
on the dipolar coupling constant $b_{14}$ (see Figure \ref{fig:1,4--mono-ammonium-maleate-spins}):
for \textbf{1}, a maximum of only $\approx16$ percent DQF efficiency
is reached for an excitation period of $\SI{7.1}{\milli\second}$,
which is considerably longer than the optimum duration of $\SI{5.7}{\milli\second}$
predicted from the value of $b_{14}/2\pi=\SI{-216}{\hertz}$. 

\begin{figure}
\begin{centering}
\includegraphics[width=0.5\columnwidth]{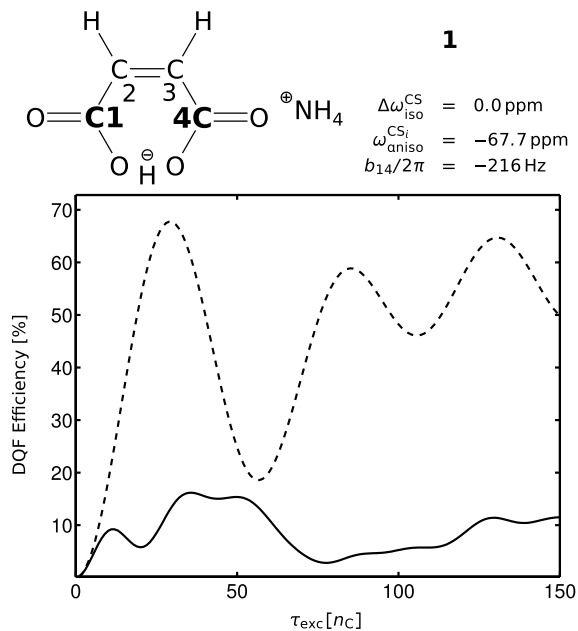} 
\par\end{centering}

\caption{\label{fig:1,4--mono-ammonium-maleate-spins}Simulated $\mathrm{C}7_{2}^{1}$
DQF buildup curves based on the known spin-system parameters \cite{Dusold1999}
of 1,4-$\ce{^{13}C2}$-mono-ammonium maleate, \textbf{1}. The lower
curve (---) includes CSA, the other curve ($\mbox{-}\mbox{-}\mbox{-}$)
assumes absence of CSA}
\end{figure}

Due to the strong effect of CSA on the DQF buildup curve (Figure \ref{fig:1,4--mono-ammonium-maleate-spins})
it is a good strategy to initially investigate the effect of every
optimisation parameter individually (durations, amplitudes, phases).
A single-parameter sensitivity test is either accomplished by step-wise
scanning the parameter over a defined range or by applying the GA
with only one variable parameter. All other parameters are fixed to
the default values of $\mathrm{C}7_{2}^{1}$. This approach also allows
to determine appropriate parameter-interval limits and bitstring
sizes (Figure \ref{fig:Binary-encoding}) for consecutive full optimisation
runs.

The most sensitive parameter under single parameter optimisation are
the pulse durations $\tau_{1}$ and $\tau_{2}$, causing an increase
in DQF efficiency from $\approx9$ percent of standard $\mathrm{C}7_{2}^{1}$
to $\approx59$ percent. This is followed by the pulse amplitudes
$\kappa_{1}$ and $\kappa_{2}$, optimisation of which increases DQF
efficiency to $\approx23$ percent. Phases $\phi_{1}$ and $\phi_{2}$
do not cause any substantial increase in DQF efficiency (compare Table
\ref{tab:GA-table}). A step-wise increase of the number of simultaneously
optimised parameters, ordered by decreasing sensitivity in single
parameter scans, provides optimum parameters as shown in Table \ref{tab:GA-table}.
\begin{table*}
\begin{centering}
\begin{tabular}{cr@{\extracolsep{0pt}.}lr@{\extracolsep{0pt}.}lr@{\extracolsep{0pt}.}lr@{\extracolsep{0pt}.}lr@{\extracolsep{0pt}.}lr@{\extracolsep{0pt}.}lcr@{\extracolsep{0pt}.}lr@{\extracolsep{0pt}.}l}
\toprule 
run & \multicolumn{2}{c}{$\tau_{1}/[\si{\micro\second}]$} & \multicolumn{2}{c}{$\tau_{2}/[\si{\micro\second}]$} & \multicolumn{2}{c}{$\kappa_{1}/[\si{\hertz}]$} & \multicolumn{2}{c}{$\kappa_{2}/[\si{\hertz}]$} & \multicolumn{2}{c}{$\phi_{1}/[\si{\degree}]$} & \multicolumn{2}{c}{$\phi_{2}/[\si{\degree}]$} & $n_{\mathrm{C}N}/[\#]$ & \multicolumn{2}{c}{$f/[0-2]$} & \multicolumn{2}{c}{$\mathrm{DQF}/[\si{\percent}]$}\tabularnewline
\midrule
\midrule 
$\mathrm{C}7_{2}^{1}$  & 14&00 & 14&00 & \multicolumn{2}{c}{71428} & \multicolumn{2}{c}{71428} & 0&00 & 0&00 & 31 & 0&91 & 8&69\tabularnewline
\midrule
GA & 14&36 & \multicolumn{2}{c}{} & \multicolumn{2}{c}{} & \multicolumn{2}{c}{} & \multicolumn{2}{c}{} & \multicolumn{2}{c}{} &  & 0&44 & 55&69\tabularnewline
GA & 13&97 & 14&27 & \multicolumn{2}{c}{} & \multicolumn{2}{c}{} & \multicolumn{2}{c}{} & \multicolumn{2}{c}{} &  & 0&41 & 59&28\tabularnewline
GA & \multicolumn{2}{c}{} & \multicolumn{2}{c}{} & \multicolumn{2}{c}{45232} & \multicolumn{2}{c}{76074} & \multicolumn{2}{c}{} & \multicolumn{2}{c}{} &  & 0&69 & 31&14\tabularnewline
GA & \multicolumn{2}{c}{} & \multicolumn{2}{c}{} & \multicolumn{2}{c}{} & \multicolumn{2}{c}{} & -82&83 & 144&28 &  & 0&87 & 13&46\tabularnewline
GA & 13&02 & 15&34 & \multicolumn{2}{c}{76643} & \multicolumn{2}{c}{66700} & \multicolumn{2}{c}{} & \multicolumn{2}{c}{} &  & 0&42 & 57&93\tabularnewline
GA & 14&05 & 14&39 & \multicolumn{2}{c}{78477} & \multicolumn{2}{c}{$=\kappa_{1}$} & \multicolumn{2}{c}{} & \multicolumn{2}{c}{} & 31 & 0&42 & 58&06\tabularnewline
GA & 12&83 & 16&02 & \multicolumn{2}{c}{77838} & \multicolumn{2}{c}{64648} & -9&94 & -6&29 &  & 0&39 & 60&80\tabularnewline
GA & 14&63 & 15&04 & \multicolumn{2}{c}{76461} & \multicolumn{2}{c}{76221} & -1&86 & -8&18 & 34 & 0&35 & 65&45\tabularnewline
\bottomrule
\end{tabular}
\par\end{centering}

\caption{\label{tab:GA-table}Optimum parameter sets from GA optimisation.
Empty cells are set to the default values of $\mathrm{C}7_{2}^{1}$;
Fitness $f$ and DQF efficiencies are best of 30 consecutive GA runs
each limited to $1500$ fitness evaluations distributed over population
sizes of $50$ individuals evolved for $30$ generations. Crossover
probability was $p_{\mathrm{c}}=0.6$ and mutation probability $p_{\mathrm{m}}=0.01$.
The GA utilised parameter boundaries centred at $\mathrm{C}7_{2}^{1}$
default values ($\Delta\tau_{i}=\pm5$ $\si{\micro\second}$, $\Delta\kappa_{i}=\pm7142$
$\si{\hertz}$, $\Delta\phi_{i}=\pm\SI{10}{\degree}$, $\Delta n_{\mathrm{C}7}=\pm20$)
and a bitstring length of $16$ for floats.}
\end{table*}
 As expected, simultaneous optimisation of all parameters results
in the overall best DQF efficiency (65.5 percent). Comparing this
to the DQF efficiency, achieved by solely optimising $\tau_{1}$ and
$\tau_{2}$, suggests that enhancements are predominantly due to changes
in pulse durations. 

\begin{figure*}
\begin{centering}
\includegraphics[width=1\textwidth]{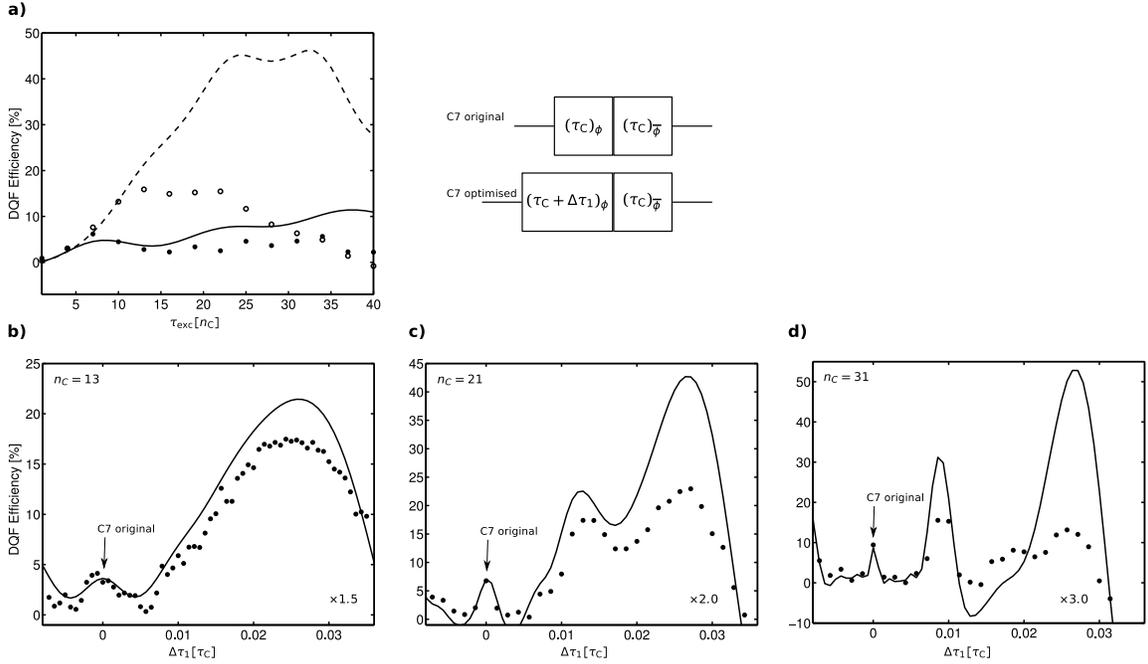}
\par\end{centering}

\caption{\label{fig:ExperimentOfOptimisedC7} DQF efficiencies of optimised
$\mathrm{C}7_{2}^{1}$ pulse sequences. a) simulated DQF buildup curves
as a function of $n_{\mathrm{C}}$ of original $\Delta\tau_{1}=\mathrm{0}$
(---) and optimised $\Delta\tau_{1}=0.026\tau_{\mathrm{C}}$ ($\mbox{-}\,\mbox{-}\,\mbox{-}$)
$\mathrm{C}7_{2}^{1}$ together with basic $\mathrm{C}$ elements
depicting the modifications. Also plotted are the respective experimental
efficiencies ($\bullet\bullet\bullet$) and ($\circ\circ\circ$).
b)--d) DQF-efficiency as a function of pulse duration change $\Delta\tau_{1}$
for different numbers $n_{\mathrm{C}}$ of $\mathrm{C}7$ blocks }
\end{figure*}

The experimental verification of DQF-efficiency enhancement by adjusting
the pulse duration $\tau_{1}=\tau_{\mathrm{C}}+\Delta\tau_{1}$ is
shown in Figure \ref{fig:ExperimentOfOptimisedC7}.
 An enhanced DQF-efficiency buildup of the optimised pulse sequence
as compared to standard $\mathrm{C7}_{2}^{1}$ can be verified experimentally
(Figure \ref{fig:ExperimentOfOptimisedC7}a). However, the agreement
between simulation and experiment is only good for excitation times
up to ca. $n_{\mathrm{C}}=21$. The numerically predicted overall
maximum DQF efficiency at $n_{\mathrm{C}}=31$ is not reached. This
degradation of experimental performance with increasing pulse sequence
length is most likely due to experimental imperfections such as accumulated
(very minor) pulse maladjustments over the large number of pulses
(868 at $n_{\mathrm{C}}=31$) and/or thermal heating of the NMR probe
caused by the continuous r.f. irradiation over an extended period
of time (e.g. $\SI{12.5}{\milli\second}$ for $n_{\mathrm{C}}=31$)
in the $\ce{^{13}C}$ channel. The change in DQF efficiency as a function
of pulse variation $\Delta\tau_{1}$ displays two clear efficiency
maxima for both the original $\mathrm{C}7_{2}^{1}$ and the optimised
sequence at $\Delta\tau_{1}=0$ and $\Delta\tau_{1}=0.026\tau_{\mathrm{C}}$
respectively (Figure \ref{fig:ExperimentOfOptimisedC7}b--d). These
two maxima occur at constant positions for different excitation times,
and the overall maximum DQF efficiency is always reached at $\Delta\tau_{1}=0.026\tau_{\mathrm{C}}$.
Both experiments and simulations generate this overall behaviour.
For shorter excitation times ($n_{\mathrm{C}}<21$) there is very
good agreement between experiments and simulations. The optimised
pulse sequence is able to experimentally achieve higher DQF efficiencies
even for non-optimum durations of the excitation/reconversion periods
(e.g. for $n_{\mathrm{C}}>21$) the efficiency is six times that of
$\mathrm{C}7_{2}^{1}$.

Next we will inspect the effect of variations of pulse durations $\tau_{1}$
and $\tau_{2}$ on DQF efficiency in somewhat more detail. Given that
these are by far the two most dominant optimisation parameters, numerical
analysis and visualisation of their effects is much facilitated as
we can safely exclude the remaining, much less sensitive parameters
from this inspection.

The symmetry rules at the core of $\mathrm{C7}_{2}^{1}$ require the
duration of a $\mathrm{C7}$ block $\tau_{\mathrm{C7}}$ to be an
integer multiple of the rotor period $\tau_{\mathrm{C7}}=7\cdot\left(2\tau_{\mathrm{C}}\right)=2\tau_{\mathrm{rot}}$.
One $\mathrm{C7}$ block duration $\tau{}_{\mathrm{C}7}$ is filled
with fourteen stacked $2\pi$ pulses of duration $\tau_{\mathrm{C}}$.
A change in the pulse duration by $\Delta\tau_{1}$ and/or $\Delta\tau_{2}$
with $\mathrm{\Delta\tau_{1}}+\Delta\tau_{2}\neq\mod2\tau_{\mathrm{C}}$
renders the $\mathrm{C7}$ block duration $\tau'_{\mathrm{C}7}=\tau{}_{\mathrm{C}7}+\Delta\tau_{\mathrm{C7}}$
asynchronous with the rotor period by 
\begin{equation}
\Delta\tau_{\mathrm{rot}}=\frac{\Delta\tau_{\mathrm{C7}}}{2}=\frac{7}{2}\left(\Delta\tau_{1}+\Delta\tau_{2}\right)\label{eq:asyncroneous}
\end{equation}
The very small optimum values of $\Delta\tau_{1}$ (Figure \ref{fig:ExperimentOfOptimisedC7})
lead to quite small deviations from synchronicity with the excitation
time $\tau_{\mathrm{exc}}=\tau_{\mathrm{C}7}n_{\mathrm{C}}$ changing
to 
\begin{equation}
\tau_{\mathrm{exc}}+\Delta\tau_{\mathrm{exc}}=\tau_{\mathrm{C}7}n_{\mathrm{C}}+\Delta\tau_{\mathrm{C}7}n_{\mathrm{C}}
\end{equation}
 Further, changing the pulse duration also alters the pulse flip angle
from $\theta_{\mathrm{C}}=2\pi$ to
\begin{eqnarray}
\theta & = & \mathrm{\theta_{\mathrm{C}}+\Delta\theta}\label{eq:flip-angle}
\end{eqnarray}

\begin{figure*}
\begin{centering}
\includegraphics[width=1\textwidth]{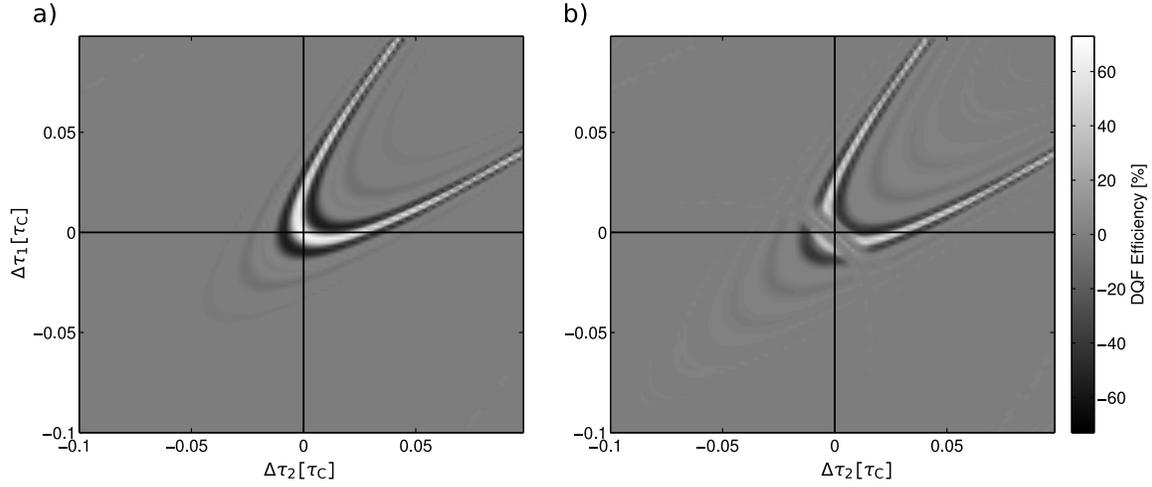}
\par\end{centering}

\caption{\label{fig:Correlation-of-Tau1Tau2}Two-dimensional contour plots
of DQF efficiency as function of $\Delta\tau_{1}$ and $\Delta\tau_{2}$;
crosshair marking indicates values of $\mathrm{C7}_{2}^{1}$. Contour
plots are based on the spin-system parameters of \textbf{1}, using
the excitation time of maximum DQF efficiency ($\tau_{\mathrm{exc}}=31\tau_{\mathrm{C}7}$)
.\textbf{ }a) Assuming absence of CSA, b) assuming presence of CSA. }
\end{figure*}

Changes of the basic $\mathrm{C}$-element pulse duration $\tau_{\mathrm{C}}$
by varying $\Delta\tau_{1}$ and $\Delta\tau_{2}$ affects the DQF
efficiency as can be seen in the contour plots in Figure \ref{fig:Correlation-of-Tau1Tau2}.
 In the absence of CSA (Figure \ref{fig:Correlation-of-Tau1Tau2}a)
a parabolically shaped area of high DQF efficiency can be identified,
where $\Delta\tau_{1}=\Delta\tau_{2}=0$ corresponds to the vertex
of the parabola and also the overall maximum DQF efficiency. This
is in perfect agreement with the predicted optimal behaviour of $\mathrm{C}7_{2}^{1}$.
In the presence of CSA (Figure \ref{fig:Correlation-of-Tau1Tau2}b),
the high DQF-efficiency parabola and a line 
\begin{equation}
\Delta\tau_{1}+\Delta\tau_{2}=0\label{eq:tau1tau2-correlation}
\end{equation}
of very low DQF efficiency occur, intersecting at the original $\mathrm{C}7_{2}^{1}$
condition. Additional local efficiency maxima occur for the conditions
$\Delta\tau_{1}+\Delta\tau_{2}=\frac{\tau_{\mathrm{C}}}{2}k+\frac{\tau_{\mathrm{C}}}{2}l$
with $k+l=4j;\ j,k,l\in\mathbb{N}$ but these are very strongly affected
by CSA (see supplementary material) 

Figure \ref{fig:Correlation-of-Tau1ExcTau2}
\begin{figure*}
\begin{centering}
\includegraphics[width=1\textwidth]{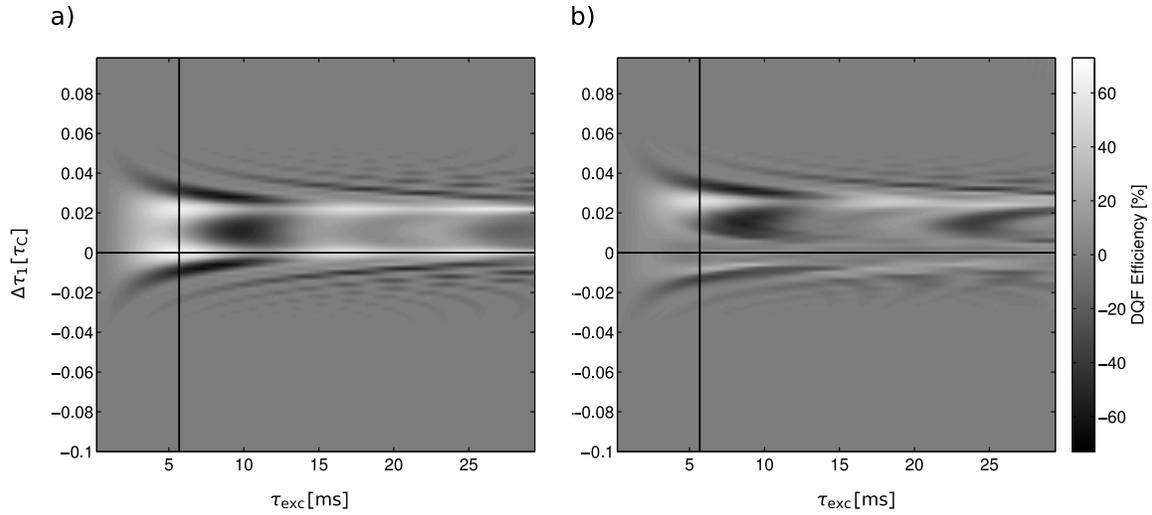}
\par\end{centering}

\caption{\label{fig:Correlation-of-Tau1ExcTau2} Two-dimensional contour plots
of DQF efficiency as function of $\Delta\tau_{1}$ and $\tau_{\mathrm{exc}}$;
crosshair marking indicates values of $\mathrm{C7}_{2}^{1}$. Contour
plots are based on the spin-system parameters of \textbf{1}.\textbf{
}a) Assuming absence of CSA, b) assuming presence of CSA.}
\end{figure*}
 displays the effect of pulse variations $\Delta\tau_{1}$ and excitation
times $\tau_{\mathrm{exc}}$ on DQF efficiency. Horizontal slices
through these contour plots represent the usual DQF-efficiency buildup
curves as are e.g. shown in Figure \ref{fig:1,4--mono-ammonium-maleate-spins}
for $\Delta\tau_{1}=0$. For vanishing CSA (Figure \ref{fig:Correlation-of-Tau1ExcTau2}a)
two symmetrically shaped horizontal bands of efficient DQF buildup
reach DQF values close to the theoretical optimum at identical buildup
rates and, therefore, reach maximum at identical excitation times
$\tau_{\mathrm{exc}}^{\mathrm{max}}$. The bands are centred at $\Delta\tau_{\mathrm{1}}=0$
and at 
\begin{equation}
\Delta\tau_{\mathrm{1}}\approx0.022\tau_{\mathrm{C}}\label{eq:optimum-tau1}
\end{equation}
This situation changes if substantial CSA is present (Figure \ref{fig:Correlation-of-Tau1ExcTau2}b).
There are still two bands present but the symmetry is broken. The
band at $\Delta\tau_{1}=0$ displays large degradation of DQF efficiency
(see Figure \ref{fig:1,4--mono-ammonium-maleate-spins}) and the (relative)
maximum DQF efficiency is reached for a different excitation time.
The second band is largely unaffected by the presence of CSA, featuring
the same behaviour as the two bands in the scenario without CSA. Hence,
for this band the DQF buildup rate is determined only by the dipolar
coupling interaction, even in the presence of substantial CSA (Vertical
slices of the contour plot, depicting this behaviour as a function
of $\Delta\tau_{1}$, are shown in Figure \ref{fig:ExperimentOfOptimisedC7}b--d).

\begin{figure*}
\begin{centering}
\includegraphics[width=1\textwidth]{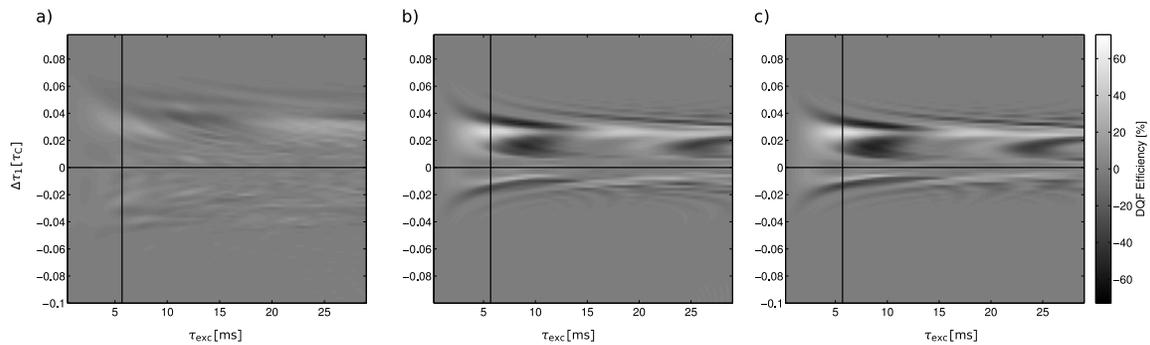}
\par\end{centering}

\caption{\label{fig:Spinning-frequency-dependence}Spinning frequency dependence
of DQF buildup of \textbf{1} as a function of excitation time \textbf{$\tau_{\mathrm{exc}}$}
and pulse duration $\Delta\tau_{1}$: a) $\omega_{\mathrm{rot}}=0.39\cdot\omega_{\mathrm{aniso}}^{\mathrm{CSA}}$,
b) $\omega_{\mathrm{rot}}=0.88\cdot\omega_{\mathrm{aniso}}^{\mathrm{CSA}}$,
c) $\omega_{\mathrm{rot}}=0.99\cdot\omega_{\mathrm{aniso}}^{\mathrm{CSA}}$
; }
\end{figure*}

In order to exploit these CSA-independent DQF-efficiency buildup conditions
it is important to be able to predict their location precisely. In
this respect the stable condition of Equation \ref{eq:optimum-tau1}
is very useful. Figure \ref{fig:Spinning-frequency-dependence} further
illustrates this stability 
for the DQF-efficiency buildup as a function of $\Delta\tau_{1}$
and $\tau_{\mathrm{exc}}$ for three different spinning speeds $\omega_{\mathrm{rot}}$
in the presence of CSA. It is apparent that for spinning speeds less
than ca. $0.9\cdot\omega_{\mathrm{aniso}}^{\mathrm{CSA}}$ the overall
DQF efficiency is fairly low. For spinning speeds equal to or larger
than the CSA interaction the desirable stable behaviour according
to Equation \ref{eq:optimum-tau1} (Table \ref{tab:Spinning-frequency-dependence})
is observed. 
\begin{table*}
\begin{centering}
\begin{tabular}{ccccc}
\toprule 
\multirow{2}{*}{$\omega_{\mathrm{rot}}/\left[\si{\hertz}\right]$} & \multicolumn{2}{c}{without CSA} & \multicolumn{2}{c}{with CSA}\tabularnewline
\cmidrule{2-5} 
 & $\tau_{\mathrm{exc}}^{\mathrm{max}}/\left[\si{\milli\second}\right]$ & $\Delta\tau_{1}^{\mathrm{max}}/\left[\tau_{\mathrm{C}}\right]$ & $\tau_{\mathrm{exc}}^{\mathrm{max}}/\left[\si{\milli\second}\right]$ & $\Delta\tau_{1}^{\mathrm{max}}/\left[\tau_{\mathrm{C}}\right]$\tabularnewline
\midrule 
$\num{4000}$ & 5.500 & 1.022 & {*}) & {*})\tabularnewline
$\num{9000}$ & 5.777 & 1.022 & 5.403 & 1.026\tabularnewline
$\num{10204}$ & 5.684 & 1.022 & 6.354 & 1.026\tabularnewline
\bottomrule
\end{tabular}
\par\end{centering}

\begin{centering}

\par\end{centering}

\caption{\label{tab:Spinning-frequency-dependence}Spinning frequency independence
of the DQF-efficiency buildup maximum for the optimised $\Delta\tau_{1}$.
Coordinates $\tau_{\mathrm{exc}}^{\mathrm{max}}$ and $\Delta\tau_{1}^{\mathrm{max}}$
are given: assuming absence of CSA and presence of CSA (compare Figure
\ref{fig:Correlation-of-Tau1ExcTau2} and \ref{fig:Spinning-frequency-dependence}).
{*}) indicates absence of a clear maximum at low spinning speeds.}
\end{table*}
 The data in Table \ref{tab:Spinning-frequency-dependence} further
confirm this robust, CSA-independent behaviour. 
\begin{figure}
\begin{centering}
\includegraphics[width=0.5\columnwidth]{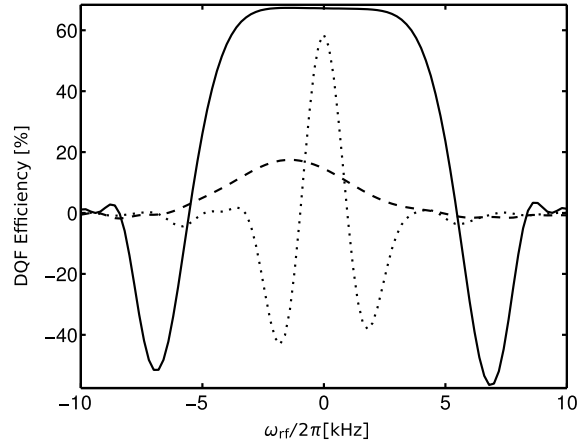}
\par\end{centering}

\caption{\label{fig:Transmitter-offset-dependence-of}Transmitter-offset dependence
of DQF efficiency for \textbf{1 }at maximum DQF efficiency ($\tau_{\mathrm{exc}}=31\tau_{\mathrm{C}7}$)
for: ideal $\mathrm{C}7_{2}^{1}$ (absence of CSA ---), original $\mathrm{C}7_{2}^{1}$
(including CSA $\mbox{-}\mbox{-}\mbox{-}$), optimised pulse sequence
parameters (including CSA $\cdot\cdot\cdot$).}
\end{figure}

Another important characteristic of recoupling MAS NMR experiments,
besides interaction-selective buildup rates, is their behaviour in
terms of broadbandedness \cite{Baldus1998}. In the absence of CSA
the $\mathrm{C}7_{2}^{1}$ pulse sequence is able to generate high
DQF efficiency over a full-width-at-half-maximum range of $\SI{9.5}{\kilo\hertz}$,
symmetrically placed around the transmitter frequency (see Figure
\ref{fig:Transmitter-offset-dependence-of}). The presence of CSA
not only reduces considerably the height of this profile, it also
breaks the symmetry of the profile around the transmitter frequency
and shifts the maximum height / maximum efficiency away from the transmitter
frequency. In practical applications, this can make it rather difficult
to choose appropriate experimental conditions. Our GA optimised pulse
sequence restores the height and symmetry of the excitation profile
around the transmitter frequency but is much more narrowbanded than
the original $\mathrm{C}7_{2}^{1}$ pulse sequence. This makes our
optimised pulse sequence a highly robust and suitable choice for selective
recoupling experiments, in some sense complementary to the original
$\mathrm{C}7_{2}^{1}$ version.

\begin{figure}
\begin{centering}
\includegraphics[width=0.5\columnwidth]{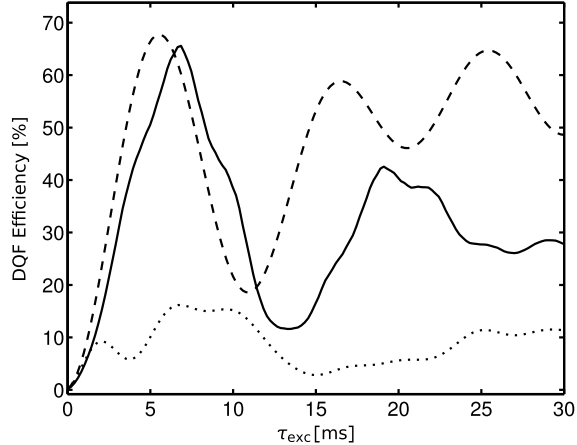}
\par\end{centering}

\caption{\label{fig:Simulated-DQF-efficiency}Simulated DQF efficiency buildup
for \textbf{1} for: an ideal $\mathrm{C}7_{2}^{1}$ buildup (absence
of CSA $\mbox{-}\mbox{-}\mbox{-}$), original $\mathrm{C}7_{2}^{1}$
buildup (including CSA $\cdot\cdot\cdot$), all pulse sequence parameters
($\kappa_{2}$, $\tau_{1}$, $\tau_{2}$, $\phi_{1}$, $\phi_{2}$,
$n_{\mathrm{C}}$) optimised (including CSA ---).}
\end{figure}

So far only the pulse durations $\tau_{1}$ and $\tau_{2}$ have been
considered in the discussion due to their high impact on the experimental
performance. Optimising all parameters ($\kappa_{1}$, $\kappa_{2}$,
$\tau_{1}$, $\tau_{2}$, $\phi_{1}$, $\phi_{2}$, $n_{\mathrm{C}}$)
results in the overall best DQF efficiency ($65.4$ percent) of the
GA optimisation (Table \ref{tab:GA-table}) and the corresponding
buildup curves are shown in Figure \ref{fig:Simulated-DQF-efficiency}.
 Comparison of the buildup curve for this best solution and the ideal
$\mathrm{C}7_{2}^{1}$ buildup (assuming absence of CSA, Figure \ref{fig:Simulated-DQF-efficiency}a)
shows very good agreement between the two. However, a slightly lower
buildup-rate for the optimised sequence is observed. This is mostly
due to the nature of the fitness function (Equation \ref{eq:fitness})
chosen such that the DQF efficiency is optimised. Improved results
might be achieved by using a multi-objective fitness function, by
parametrising DQF efficiency and buildup rate together. The fully
optimised solution shows the very high robustness of the fully optimised
solution to different magnitudes of CSA interactions. However, it
also is found that the fully optimised solution has a even more narrowbanded
profile of its recoupling performance as compared to the optimisation
of $\tau_{1}$ alone. This narrowbandedness reflects the fact that
the used fitness function does not penalise a narrowbanded characteristic
of the resulting pulse sequence.

\subsection{Comparison of GA with other Algorithms}

Due to their stochastic nature, GAs have many traits of sampling the
search space at random points, or of complete sampling of the search
space at a given grid size. The optimum parameters shown in Table
\ref{tab:GA-table} are each the best result of $30$ GA runs, with
each run limited to $\num{1500}$ fitness function evaluations. The
same optimisation is now iterated for $\num{1500}$ randomly sampled
points in search space for $30$ runs (see Table \ref{tab:Parameters-randomly-sampled}).
Comparison of the resulting DQF efficiencies shows that for small
numbers of optimisation parameters there is no difference between
random sampling and GA performance, given that 1500 points probe the
search space sufficiently accurately. However, with increasing numbers
of parameters the probing gets coarser and the deviation between random
sampling and GA best result increases. One might argue that the stochastic
nature of the two approaches makes the best result of $30$ runs possibly
merely a matter of luck. A well behaved algorithm, however, should
also display a good chance to reproduce a successful search at a reasonable
rate. This can be assessed e.g. by considering every run finding a
DQF efficiency larger than 50 percent a success. For the scenario
where all parameters are optimised simultaneously the GA has a 50
percent chance of success while random sampling only has a 17 percent
chance of success.
\begin{table*}
\begin{centering}
\begin{tabular}{cr@{\extracolsep{0pt}.}lr@{\extracolsep{0pt}.}lr@{\extracolsep{0pt}.}lr@{\extracolsep{0pt}.}lr@{\extracolsep{0pt}.}lr@{\extracolsep{0pt}.}lcr@{\extracolsep{0pt}.}lr@{\extracolsep{0pt}.}l}
\toprule 
run & \multicolumn{2}{c}{$\tau_{1}/[\si{\micro\second}]$} & \multicolumn{2}{c}{$\tau_{2}/[\si{\micro\second}]$} & \multicolumn{2}{c}{$\kappa_{1}/[\si{\hertz}]$} & \multicolumn{2}{c}{$\kappa_{2}/[\si{\hertz}]$} & \multicolumn{2}{c}{$\phi_{1}/[\si{\degree}]$} & \multicolumn{2}{c}{$\phi_{2}/[\si{\degree}]$} & $n_{\mathrm{C}N}/[\#]$ & \multicolumn{2}{c}{$f/[0-2]$} & \multicolumn{2}{c}{$\mathrm{DQF}/[\si{\percent}]$}\tabularnewline
\midrule
\midrule 
$\mathrm{C}7_{2}^{1}$  & 14&00 & 14&00 & \multicolumn{2}{c}{71428} & \multicolumn{2}{c}{71428} & 0&00 & 0&00 & 31 & 0&91 & 8&69\tabularnewline
\midrule
random & 14&36 & \multicolumn{2}{c}{} & \multicolumn{2}{c}{} & \multicolumn{2}{c}{} & \multicolumn{2}{c}{} & \multicolumn{2}{c}{} &  & 0&44 & 55&69\tabularnewline
random & 13&97 & 14&27 & \multicolumn{2}{c}{} & \multicolumn{2}{c}{} & \multicolumn{2}{c}{} & \multicolumn{2}{c}{} &  & 0&42 & 58&38\tabularnewline
random & \multicolumn{2}{c}{} & \multicolumn{2}{c}{} & \multicolumn{2}{c}{44323} & \multicolumn{2}{c}{75884} & \multicolumn{2}{c}{} & \multicolumn{2}{c}{} &  & 0&69 & 30&90\tabularnewline
random & \multicolumn{2}{c}{} & \multicolumn{2}{c}{} & \multicolumn{2}{c}{} & \multicolumn{2}{c}{} & -7&95 & 147&37 &  & 0&87 & 13&47\tabularnewline
random & 13&88 & 14&38 & \multicolumn{2}{c}{74524} & \multicolumn{2}{c}{70358} & \multicolumn{2}{c}{} & \multicolumn{2}{c}{} &  & 0&43 & 56&64\tabularnewline
random & 14&46 & 14&02 & \multicolumn{2}{c}{66234} & \multicolumn{2}{c}{$=\kappa_{1}$} & \multicolumn{2}{c}{} & \multicolumn{2}{c}{} & 32 & 0&47 & 52&59\tabularnewline
random & 14&87 & 13&38 & \multicolumn{2}{c}{64452} & \multicolumn{2}{c}{72880} & 9&70 & -2&67 &  & 0&43 & 56&87\tabularnewline
random & 14&62 & 14&83 & \multicolumn{2}{c}{78208} & \multicolumn{2}{c}{75878} & 1&26 & -4&99 & 38 & 0&38 & 62&29\tabularnewline
Simplex & 14&61 & 15&01 & \multicolumn{2}{c}{76504} & \multicolumn{2}{c}{76215} & -1&86 & -8&17 & 34 & 0&34 & 65&73\tabularnewline
\bottomrule
\end{tabular}
\par\end{centering}

\caption{\label{tab:Parameters-randomly-sampled}Optimum parameter sets from
random sampling. Empty cells correspond to values of $\mathrm{C}7_{2}^{1}$;
Values are best of 30 repeated GA runs; each comprising 1500 fitness
evaluations, spread over population sizes of 50 individuals evolved for
30 generations.
Crossover probability was $p_{\mathrm{c}}=0.6$ and mutation probability
$p_{\mathrm{m}}=0.01$. Random sampling utilised parameter boundaries
centred at $\mathrm{C}7_{2}^{1}$ default parameters ($\Delta\tau_{i}=\pm5$
$\si{\micro\second}$, $\Delta\kappa_{i}=\pm7142$ $\si{\hertz}$,
$\Delta\phi_{i}=\SI{\pm10}{\degree}$, $\Delta n_{\mathrm{C}7}=\pm20$) }
\end{table*}

For comparison of the GA to non-heuristic search algorithms MIGRAD
\cite{James1975} and Simplex \cite{PFT+92} are chosen, representing
the classes of gradient and non-gradient based search algorithms.
Heuristic (GA) searches gain from large numbers of fitness-function
evaluations due to the concomitant increase in population and generation
sizes. Non-heuristic methods are affected by a fitness-function evaluation
limit to a much lesser extend. These algorithms strongly dependent
on a good initial guess of starting parameter values and on the structure
of the fitness-function landscape. More specifically, the Simplex
algorithm is mostly affected by the existence of local maxima in the
vicinity of its starting point, but is quite independent of the shape
of the maximum. The MIGRAD algorithm, like most gradient based algorithms,
assumes that the maximum region can be approximated by a quadratic
function and hence takes advantage of the definition of the fitness
function and, in turn, tends to converge faster. MIGRAD and Simplex
are unconstrained and take confidence parameters for every parameter
instead. These control the variation of the optimisation parameters
in the initial steps of the algorithm.

The error landscapes displayed in Figure \ref{fig:ExperimentOfOptimisedC7}b--d
and Figure \ref{fig:Correlation-of-Tau1Tau2} can be considered as
typical for NMR pulse sequence optimisation, especially for those
using resonance conditions to achieve re/decoupling \cite{Scholz2010}.
Choosing confidence parameters identical to the parameter interval
limits of the constraint GA optimisation and using the original $\mathrm{C}7_{2}^{1}$
parameters as initial parameters, Simplex manages to evade the local
maximum within the limit of $\num{1500}$ steps, while MIGRAD fails
to do so. Starting with initial parameters further away from the optimum
causes both algorithms to struggle even more in achieving good efficiencies.

As a test of the quality of the optimum parameters found by the GA,
these can be passed on to additional Simplex optimisation (compare
bottom of Table \ref{tab:GA-table} and \ref{tab:Parameters-randomly-sampled}).
Only a very slight further improvement (0.3 percent) is found. This
can be ascribed just to the now continuous parameters as opposed to
the discrete values of the GA parameters resulting from the bit-depth
of encoding. Combinations of genetic algorithms and classical optimisation
has been successfully exploited in other areas \cite{Waechtler2009,Ong2006,Hibbert1993}
for the refinement of search results but this does not seem to offer
great advantages when searching for improved NMR pulse sequences.
In some circumstances GA optimisations may be computationally dearer than classical algorithms
but the reward may very well be the ability to find solutions not accessible to 
other search algorithms.

Optimal control approaches have featured prominently in recent years
in searches for improved NMR pulse sequences \cite{Tosner2009,Nielsen2007}.
Various methods of optimal control (dynamic programming, Pontryagin
minimum principle \cite{Kirk2004}) have been applied successfully
using both geometrical and numerical approaches. The numerical approaches
(specifically the GRAPE or Krotov implementations of optimal control
\cite{Khaneja2005,Maximov2008}) and GA algorithms have in common
that both can be applied to some extent in a black-box fashion. This
makes both methods attractive tools for general applications. Apart
from this, GA and optimal control have to be regarded rather complementary
in their respective strengths. Due to the dependence of optimal control
algorithms on gradient information they are affected by the shape
of search landscapes and initial parameter values in quite the same
way as are classical gradient based algorithms, where the presence
of local minima may pose problems. GAs are better placed to handle
such circumstances successfully. If the search space is rather smooth
and optimum conditions can be reached with a continuous increase of
pulse sequence performance from nearly every initial condition it
may well be the case that optimal control approaches converge faster
than GA searches. In searches where the optimum condition is a fairly
sharp, singular resonance condition (such as is common for recoupling
MAS NMR experiments) GA searches may be advantageous. GAs and optimal
control approaches, in principle, have in common that both may produce
results that make physical insight difficult. As far as optimal control
is concerned, the need to simplify and clarify optimisation results
has been recently addressed by introducing smoothing constraints to
successive pulses \cite{Maximov2010} and modularisation of the pulse
sequence \cite{Nielsen2009} in order to aide the generation of structured
results that later can be rationalised by theory. Guiding GA searches
by biasing the algorithms toward certain structured results is certainly
also possible. In other areas of application for optimisation procedures,
it has been suggested to combine GA and optimal control algorithms
for situations where the structure of the search space can not be
handled well by only one of the two approaches \cite{Roy2009,Park2004,Zeidler2001,Amstrup1995,Michalewicz1992}.

\section{Experimental}

$\ce{^{31}C}$ NMR spectra spectra were recorded using a Bruker Avance
II 700 $\left(\omega_{0}/2\pi=\SI{-176.1}{\mega\hertz}\right)$ spectrometer
equipped with a $\SI{2.5}{\milli\meter}$ TriGamma MAS probe. MAS
spinning speed was $\SI{10204}{\hertz}$ throughout. Heteronuclear
$\ce{^{1}H}$ decoupling was suspended during $\mathrm{C7}$ irradiation
\cite{Hughes2004,Marin-Montesinos2005}, during acquisition $\SI{71.4}{\kilo\hertz}$
$\ce{^{1}H}$ cw decoupling was applied. Hartmann-Hahn cross-polarisation (CP)
contact time was $\SI{5}{\milli\second}$. $32$ scans were accumulated
per spectrum with $\SI{20}{\second}$ recycle delays. 
$2\pi$-pulse r.f. amplitudes  were calibrated on \textbf{1}. After CP, the $\ce{^{13}C}$ 
signal of  a train of  $2\pi$-pulses is maximised (comparing 1, 7 and 15 pulses).
Different phase cycling schemes did not provide improvements over a basic 32-step cycle.
Active temperature control of the sample ($T=\SI{298}{\kelvin}$) was used throughout.
1,4-$\ce{^{13}C2}$-mono-ammonium maleate was co-crystallised \cite{Dusold1999} with mono-ammonium
maleate ($\ce{^{13}C}$ nat. abund.) in ratios 1:7 and 1:15 to suppress
inter-molecular dipolar couplings effects on  spin-pair behaviour
(1:7 dilution found to be sufficient). Simulations used the SIMPSON
\cite{Bak2000} package. The genetic algorithm was implemented as
an extension to SIMPSON's own fit function using the GAlib \cite{Wall1996}
library. The Simplex and MIGRAD algorithms were used in their SIMPSON
implementation.

\section{Conclusions and Outlook}

GAs offer powerful and flexible tools for the optimisation of NMR
pulse sequences. The most important step in such procedures is to
derive a suitable encoding of the pulse sequence parametrisation as
a genome, operated on by the algorithm. Since GAs offer huge flexibility
of exposing specific NMR characteristics to the algorithm, this step
requires careful choices in order to exploit the properties of GAs
to their full capacity. In our example, it would not have been a clever
choice to expose all pulse sequence parameters simultaneously and
to equal extent as this would have led to a very large search space
and impractical computational cost. However, GAs do allow one to make
task-specific choices that, for example, draw from existing theoretical
knowledge about C-sequences while allowing sufficient degrees of freedom
to find optimised solutions outside the boundaries of the initial
theoretical design principles. Our optimised solution maintains by
and large the seven-fold symmetry principle of the sequence, it only
slightly breaks the synchronicity of the pulse sequence with the rotation
period of the sample. 
Wether or not, and if so, to which extent, the  cyclic nature of the pulse sequence 
is affected  depends on the spin-system properties and will differ 
for cases with and without CSA. Given that our optimisations yield performance improvements 
for both cases, one may cautiously speculate that the cyclic nature of the sequence is
less important.
This finding should represent a promising starting
point for further theoretical investigations into non-synchronous
recoupling MAS NMR pulse sequences. To the best of our knowledge,
this has not yet been explored. We note in passing that the occurrence
of small deviations from perfect symmetry having major impacts is
quite common in the physical world, ranging, for example, from so-called
incommensurate structures in crystallography \cite{VanSmallen1995}
to the properties of viruses \cite{Keef2008}, or the effects of
minor decoherences in quantum information \cite{Kendon2003}.

Open source NMR simulation software and GA programming libraries make
it technically straightforward to assemble task-specific optimisation
routines, far more sophisticated than the example we have discussed
here. For example, multi-objective fitness functions are promising
candidates for highly specific search tasks. In addition, more advanced
genomic structures and genetic operators may be exploited in general.
Rather specific to GA searches in NMR are the uniquely well structured
characteristics of the underlying Hamiltonians describing the spin
dynamics. Such a search environment may benefit hugely from GA approaches
such as Grammatical Evolution and Cartesian Genetic Programming \cite{Poli2008,ONeill2003,Miller2000}
where the algorithm itself is well structured. This, in turn, may
lead not only to improved performance of pulse sequences but also
to improved insight into the optimised NMR experiments.

\section{Acknowledgements}

Support of this work by the Deutsche Forschungsgemeinschaft, the Royal
Society (JP080770),  EPSRC grant EP/D050618/1 (SEBASE) and EP/J017515/1
(DAASE) are gratefully acknowledged. We thank the NMR Centre at Warwick
University for spectrometer access, enabling preliminary experiments.
We thank TIFR, Mumbai, and P.\ K. Madhu for hosting us and for making
available generous laboratory access.

\bibliographystyle{plain}
\bibliography{JMR-12-297Rev}

\end{document}